\documentclass[twocolumn,showpacs,preprintnumbers,amsmath,amssymb,prb]{revtex4}

\usepackage{graphicx}
\usepackage{dcolumn}
\usepackage{bm}

\begin{document}

\title{Modification of Born impurity scattering near the surface of
$d$-wave superconductors and influence of external magnetic field}

\author{A.~Zare, A.~Markowsky, T.~Dahm, and N.~Schopohl}

\affiliation{Institut f\"ur Theoretische Physik and Center for 
Collective Quantum Phenomena, Universit\"at T\"ubingen,\\
             Auf der Morgenstelle 14, D-72076 T\"ubingen, Germany}

\date{\today}

\begin{abstract}
We study the influence of self-consistent Born impurity 
scattering on the
zero-energy Andreev bound states near the surface of a
$d$-wave superconductor with and without an externally
applied magnetic field. Without an external magnetic field
we show that the effect of Born impurity scattering is
stronger at the surface than in the bulk. 
In the presence
of an external magnetic field the splitting of the
zero-energy Andreev bound states is shown
to have a nonmonotonous temperature dependence.
Born impurity scattering does not wash out the peak splitting, 
but instead the peak splitting is shown to be quite robust against
impurities. We also show that a nonzero renormalization of
the pair potential appears near the surface.
\end{abstract}


\maketitle

\section{\label{sec:intro} Introduction}

At the surface of $d$-wave superconductors zero-energy Andreev bound states
may appear depending on the orientation of the $d$-wave with respect to
the surface normal \cite{Hu,Tanaka95,Buchholtz,KashiwayaReport}.
Experimentally, these states can be observed as zero-bias conductance
peaks in the tunneling conductance \cite{Covington,Chesca,Deutscher,Wagenknecht}. 
It is well known that surface roughness, surface disorder \cite{Kalenkov},
or diffuse scattering at the surface leads to a broadening of these states.
Also, impurity scattering in the bulk of the superconductor is known
to broaden the Andreev bound states \cite{Poenicke,Tanaka}.
In the presence of an external magnetic field the screening current
leads to a splitting of the Andreev bound states \cite{Fogelstroem,Aprili,Krupke}.
In this case, a counter-flowing para\-magnetic current is generated by the surface states, 
which increases with decreasing temperature and may even lead to
a reversal of the current flow at the surface resulting in an
anomalous Meissner effect \cite{Fogelstroem,Walter}. This effect has
recently been shown to have a strong influence on the Bean-Livingston
surface barrier for entrance of vortices into the superconductor \cite{Iniotakis}.

In the present work we investigate the influence of
bulk impurity scattering on the broadening of the surface Andreev bound 
states and the splitting in an external magnetic field. It has been
shown previously that impurity scattering in the Born limit is much
more effective in broadening the Andreev bound states than
impurity scattering in the unitarity limit \cite{Poenicke,Tanaka}.
In the high-$T_c$ cuprate compounds it is believed that scatterers
within the CuO$_2$ planes act as unitary scatterers and thus should
have little influence on the Andreev bound states. However, it
has been recognized recently that scatterers sitting between the 
CuO$_2$ planes are poorly screened and act as Born scatterers 
\cite{Abrahams,Zhu,Dahm}. These impurities are thus expected to have 
a dominating influence on the broadening of the Andreev bound states. 
For these reasons in the present work we will focus on the influence of 
impurity scattering in the self-consistent Born approximation. We will 
show that in this case
impurity scattering around zero energy is significantly increased 
near the surface as compared to the bulk, leading to a larger
broadening of the Andreev bound states than expected from the
scattering rate in the bulk. The situation changes completely in 
the presence of an external magnetic field, however. The splitting
of the Andreev bound states turns out to be quite robust against
Born impurity scattering.

In the bulk of a $d$-wave superconductor the renormalization
of the pair potential due to impurity scattering is known to disappear. 
However, this is not generally the case near a surface because 
of broken translational symmetry. Here, we will show that a 
nonzero renormalization of the pair potential appears near the surface unless 
the orientation of the surface is highly symmetric with
respect to the orientation of the $d$-wave. Also, in the
presence of an external magnetic field the
renormalization of the pair potential becomes nonzero.

In Section \ref{sec:1} we will describe our numerical
approach. In Section \ref{sec:2} we will first study
impurity scattering near a surface of a $d$-wave
superconductor in the absence of an external magnetic field.
Section \ref{sec:3} presents results for a superconductor 
without impurities in the presence of an external magnetic field,
and Section \ref{sec:4} deals with both impurity scattering
and the presence of an external magnetic field.

\section{\label{sec:1} Numerical approach}

\begin{figure}[t]
  \begin{center}
\includegraphics[width=0.45\textwidth]{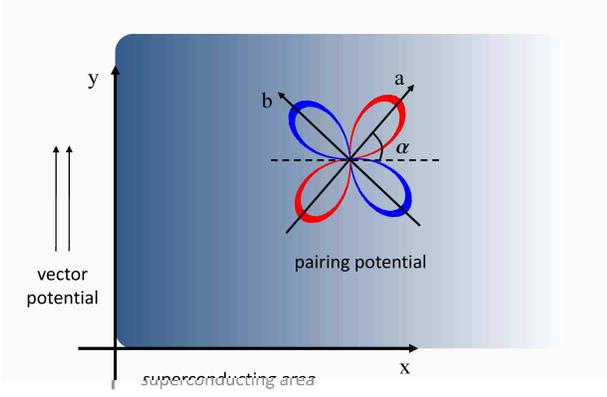}
\caption{(Color online) \label{geometry} Sketch of the investigated geometry, 
showing the pair potential and the direction of the vector potential relative 
to the surface of the superconductor. The angle $\alpha$ determines the relative
orientation of the $d$-wave with respect to the surface normal.}
  \end{center}
\end{figure} 

The geometry under investigation is shown in Fig.~\ref{geometry}: in the
halfspace $x > 0$, we assume to have a superconducting area of $d$-wave type. 
For $x < 0$, an external magnetic field $\vec B = B \vec e_{z}$ is applied
parallel to the z-axis.  For simplicity, we consider a cylindrical 
Fermi surface with the c-axis oriented parallel to the z-axis. Also,
we assume that the external magnetic field remains smaller than the
field of first vortex penetration. Therefore, we can assume 
translational invariance both along the y-axis as well as along the z-direction. 
The relation between the current density in the superconductor and the 
vector potential in Coulomb gauge is given by Maxwell's equation:

\begin{eqnarray}\label{equ-maxwell}
 - \Delta \vec A = \frac{4\pi}{c} \vec j
\end{eqnarray}
In this gauge the vector potential is directly proportional to the
superfluid velocity $v_s$.
The boundary conditions for this second order differential equation are 
determined by the behavior of the magnetic field: it penetrates into 
the superconductor continuously and decays to zero in the bulk.
 
Our calculations are based on the Eilenberger equations \cite{Eilenberger,Larkin}. 
These equations can be transformed into Riccati type differential equations 
for two scalar complex quantities $a(s)$ and $b(s)$ along real
space trajectories $\vec{R}(s)=\vec{r}+ s \hat{v}_F$ \cite{Schopohl}: 

\begin{eqnarray}
\hbar v_{F}\frac{\partial}{\partial s} a(s) + [2 \tilde \epsilon_{n}
(s) + \tilde \Delta^{\dagger}(s) a(s)]a(s) -\tilde \Delta (s) &=& 0 \nonumber \\ 
\hbar v_{F}\frac{\partial}{\partial s} b(s) - [2 \tilde \epsilon_{n}(s) 
+ \tilde \Delta(s) b(s)]b(s) + \tilde \Delta^{\dagger}(s) &=& 0
\label{Riccatieqs}
\end{eqnarray}
Here, $v_{F}$ is the Fermi velocity and $\hat{v}_F$ the unit vector in the
direction of the Fermi velocity.
The initial values for solving the Riccati equations for $a(s)$ and $b(s)$ 
are obtained from the fact, that their variation vanishes in the bulk:
\begin{eqnarray}
a(-\infty)= \frac{\tilde \Delta(-\infty)}{\tilde \epsilon_{n}+ \sqrt{\tilde \epsilon_{n}^2
+ |\tilde \Delta(-\infty)|^2}}
\end{eqnarray}
\begin{eqnarray}
b(+\infty)= \frac{\tilde \Delta^{\dagger}(+\infty)}{\tilde \epsilon_{n}+
\sqrt{\tilde \epsilon_{n}^2 + |\tilde \Delta(+\infty)|^2}}
\end{eqnarray}
Here, the renormalized Matsubara energies and pair potential are given by:
\begin{eqnarray*}
i\tilde{\varepsilon}_n (\vec{R}(s),\varepsilon_n ,T) &=& i\varepsilon_n+
\frac{e}{c}\vec{v}_F \cdot \vec{A}(\vec{R}(s))-\Sigma^G(\vec{R}(s),\varepsilon_n ,T)
\\
\tilde{\Delta}(\vec{R}(s),\varepsilon_n,T) &=& \Delta(\vec{R}(s),T)
\cos(2(\theta - \alpha))+ \\
&& \Sigma^F(\vec{R}(s),\varepsilon_n ,T)
\end{eqnarray*}
where $\Sigma^{G}$ and $\Sigma^{F}$ denote the diagonal and off-diagonal self energies
due to impurity scattering.
The pair potential is given by:
\begin{eqnarray}\label{math-num-ordnung}
\Delta(\vec{r}, T) = V N_0 \pi T  \sum_{\vert \epsilon_{n} \vert < \omega_{c} }
\langle \cos(2(\theta - \alpha)) f(\vec{r}, \vec k_F, i \epsilon_{n})
\rangle_{FS}
\end{eqnarray}
The brackets $\langle \cdots \rangle_{FS}$ denote an angular average over the 
cylindrical Fermi-surface.
By solving the Riccati equations along real space trajectories $\vec R(s)$ 
running parallel to the Fermi velocity $\vec v_{F}$, the normal and anomalous 
propagators are found from:
\begin{eqnarray}
g(\vec{R}(s)) = (-i)\cdot\frac{1-a(s)b(s)}{1+a(s)b(s)},\quad  f(\vec{R}(s)) = 
\frac{2a(s)}{1+a(s)b(s)}
\label{propagators}
\end{eqnarray}
From the propagators, we can instantly derive the current density
$\vec j (\vec r)$ and the local density of states (LDOS) $N(E, \vec{r})$:
\begin{eqnarray}\label{math-ldos}
\frac{N(E,\vec{r})}{N_0} = - {\mathrm{Im}} \, \langle g(\vec{r},\vec{k}_{F},
i \epsilon_{n} \rightarrow E + i0^+)  \rangle_{FS}
\end{eqnarray}
where $N_0$ is the normal state density of states at the Fermi level, and

\begin{eqnarray}\label{math-num-strom}
\vec j(\vec{r},T) = 2 e N_0 v_{F} k_{B}\pi T  \sum_{\vert \epsilon_{n} \vert <
\omega_{c} } \langle \hat v_{F}\cdot g(\vec{r}, \vec k_F, i \epsilon_{n})
\rangle_{FS}
\end{eqnarray}
The zero temperature London penetration depth $\lambda_L$ is given by the expression
$\lambda_L^{-2}=\frac{4\pi}{c^2}e^2 N_0 v_F^2$. Throughout the work
we will quote $\lambda_L$ relative to the zero temperature coherence 
length without impurities $\xi_{0}=\frac{\hbar v_F}{\pi\Delta(T=0)}$, i.e.
the parameter $\kappa=\lambda_L/\xi_{0}$. Magnetic fields will be given
in units of the zero temperature upper critical field $B_{c2}$, which
for a bulk $d$-wave superconductor is given by 
$B_{c2}=0.49 \frac{\Phi_0}{2\pi\xi_0^2}$ for a cylindrical
Fermi surface.

In our model, we include the effect of impurity scattering in the 
self-consistent Born approximation. 
As pointed out above, Born impurity scattering is expected to cause a
stronger effect than
scattering in the unitarity limit \cite{Poenicke,Tanaka}. For simplicity,
we will restrict ourselves to $s$-wave scattering. In this case 
the impurity self energies are given by:
\begin{eqnarray}
\Sigma^F(\vec{r}, \epsilon_n ,T) &=& \frac{1}{2\tau}\langle 
f(\vec{r},\vec{k}_F,\epsilon_n)\rangle_{FS} \label{selfenergiesF}
\\
\Sigma^G(\vec{r},\epsilon_n ,T) &=& \frac{1}{2\tau}\langle 
g(\vec{r},\vec{k}_F,\epsilon_n)\rangle_{FS}
\label{selfenergies}
\end{eqnarray}
where $\tau$ is the scattering lifetime in the bulk and is given by:
\begin{eqnarray*}
\frac{1}{\tau}=\frac{2}{\hbar}\pi N_0 n_i \vert V_0\vert ^2
\end{eqnarray*}
where $n_i$ is the impurity concentration, $V_0$ is the strength of 
the impurity potential.
Throughout this work, we will quantify the impurity scattering in terms of its 
mean free path $l=v_F\tau$ relative to the zero temperature bulk coherence 
length in the superclean limit $\xi_{0}$.
In the bulk of a $d$-wave superconductor the angular average of the anomalous
Green's function $f$ over the
Fermi surface vanishes, because positive and negative contributions cancel
exactly. This leads to a vanishing renormalization $\Sigma^F$
of the pair potential.
However, as we will show below, this does not generally hold anymore 
in the vicinity of a surface.

\begin{figure}
\begin{center}
\includegraphics[width=0.45\textwidth]{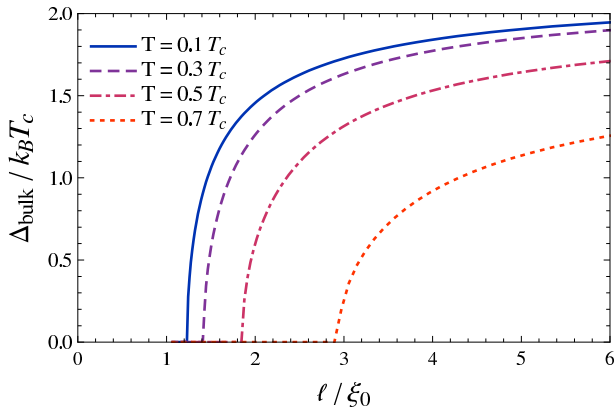}
\caption{\label{Gapbulk} (Color online) Bulk value of the pair
potential as a function 
of mean free path $l/\xi_0$ for different temperatures.}
\end{center}
\end{figure}

Numerically, we start with estimated functions
$\tilde{\Delta}(x)$ and $\vec{A}(x)$. These are used to solve the Riccati
equations (\ref{Riccatieqs}) along all real space trajectories with
specular reflection on the surface $x=0$. From the solutions we find
the propagators Eq.~(\ref{propagators}). These are used to obtain the
self energies Eq.~(\ref{selfenergiesF}) and (\ref{selfenergies}), the current density 
Eq.~(\ref{math-num-strom}), and the updated pair potential 
Eq.~(\ref{math-num-ordnung}). Finally, integration of Eq.~(\ref{equ-maxwell})
yields an updated vector potential. This procedure is iterated
until the functions $\tilde{\Delta}(x)$ and $\vec{A}(x)$ converge.
Note, that the self energies $\Sigma^G$ and $\Sigma^F$ and the propagators
$g$ and $f$ are calculated self-consistently this way.
After convergence, a final iteration is run, in which all equations
are solved directly for real frequencies $i \epsilon_{n} \rightarrow E + i0^+$
in order to perform an analytic continuation for the local density
of states and the self energies. Only for the calculations without
impurities we have added a small imaginary part of 
$0.007 k_B T_c$ in order to regularize the poles of the propagators. 

\section{\label{sec:2} Impurity scattering near a surface of a $d$-wave
superconductor}

In this section we consider a superconductor with different impurity 
concentrations for the case that no external magnetic field is applied. 
It is well known that impurities in bulk $d$-wave superconductors lead 
to pair breaking, which implies a decrease of the bulk order 
parameter. Fig.~\ref{Gapbulk} shows the bulk value of the pair
potential as a function of the mean free path $l$ for different temperatures. 
When the mean free path becomes comparable to the finite temperature
coherence length, the pair
potential vanishes. Consequently, for $d$-wave superconductors, 
there exists no real dirty limit.

Near the surface Andreev bound states are absent for $d$-wave orientation
$\alpha=0$. When the angle $\alpha$ is increased, the spectral weight
of the Andreev bound states gradually increases until it reaches a
maximum at $\alpha=\pi/4$. In the following we will first focus on an
intermediate angle of $\alpha=\pi/8$, where the spectral weight is
neither absent nor fully developed.

\begin{figure}
\begin{center}
\includegraphics[width=0.45\textwidth]{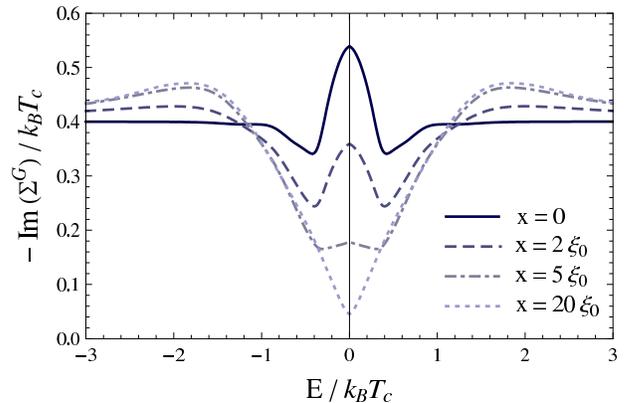}
\caption{\label{Im-Self-G-vs-E-alpha-pi-8} (Color online) 
Negative imaginary part of self energy $\Sigma^{G}$ 
as a function of energy $E$  for different distances from the surface at an 
orientation angle of $\alpha = \frac{\pi}{8}$. The temperature is $T = 0.1 T_{c}$ 
and the mean free path $l = 2.7 \xi_{0}$.}
\end{center}
\end{figure}

The local quasiparticle scattering rate is given by the negative imaginary 
part of the normal self energy -Im~$\Sigma^G$. In 
Fig.~\ref{Im-Self-G-vs-E-alpha-pi-8} we show the energy dependence of
-Im~$\Sigma^G$ for different distances from the surface at a temperature
$T=0.1 T_c$ and a mean free path of $l = 2.7 \xi_0$. From this it can
be seen that there is a significant variation of the quasiparticle scattering 
rate at the Fermi level $E=0$ as a function of the distance from the surface.
For this set of parameters, at the surface the quasiparticle
scattering rate is about 12 times larger
than in the bulk. Physically, this effect can be understood
from Eq.~(\ref{selfenergies}). At the surface in the presence of the
Andreev bound states there is a larger phase space of low energy states 
available for scattering. This makes impurity scattering more 
effective at the surface than in the bulk.

\begin{figure}[t]
\begin{center}
\includegraphics[width=0.45\textwidth]{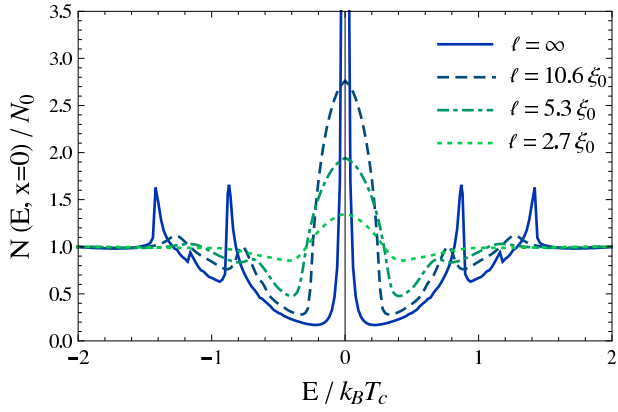}
\caption{\label{Dos-wall-pi-8-impurity-B-0-T-01}  (Color online)
Local density of states at 
the surface for different impurity concentrations. The temperature is 
$ T=0.1 T_c$ and the orientation angle $\alpha=\frac{\pi}{8}$. }
\end{center}
\end{figure}

In Fig.~\ref{Dos-wall-pi-8-impurity-B-0-T-01}, the impurity dependence of the 
local density of states at the surface is shown for the same angle $\alpha$ and
temperature. Increasing the impurity concentration results in a decrease of the 
height of the zero energy peak and a broadening of its width. The peaks
seen near $1.4 k_B T_c$ and $0.9 k_B T_c$ in the absence of impurity scattering
can be interpreted as follows: the peaks near $\pm 1.4 k_B T_c$ are related
to the bulk gap times $\cos 2\alpha$. They are coming from quasiparticles, which
approach the surface perpendicular, as a look at the momentum resolved
data shows. In contrast, the peaks near $\pm 0.9 k_B T_c$ are
caused by grazing angle quasiparticles. In our self-consistent
calculation the local gap near the surface is much smaller
than in the bulk. The grazing angle quasiparticles mostly
experience the reduced surface gap value, creating a
gap edge around $0.9 k_B T_c$. In the presence of impurities these
gap features are quickly washed out.

\begin{figure}[t]
\begin{center}
\includegraphics[width=0.45\textwidth]{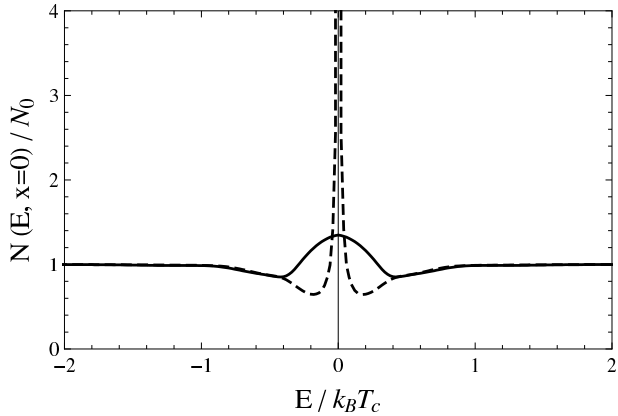}
\caption{\label{scattering-stength} The local density of states at the surface 
for temperature $T=0.1 T_c$, orientation angle $\alpha = \frac{\pi}{8}$ and 
mean free path $l=2.7 \xi_{0}$. The dashed curve shows the LDOS when the bulk 
values of the self energy are used. The solid curve, on the other hand, shows 
the LDOS when using the self consistent solution for the self energy.}
\end{center}
\end{figure}

In order to illustrate the change of the local density of states at the surface
due to impurity scattering, in Fig.~\ref{scattering-stength} we compare
the local density of states at the surface for $l = 2.7 \xi_0$ with a hypothetical
calculation, in which we have used the bulk value of $\Sigma^G$ at the surface.
Clearly, the zero energy peak is much sharper when the bulk  $\Sigma^G$
is used for the calculation. 

\begin{figure}
\begin{center}
\includegraphics[width=0.45\textwidth]{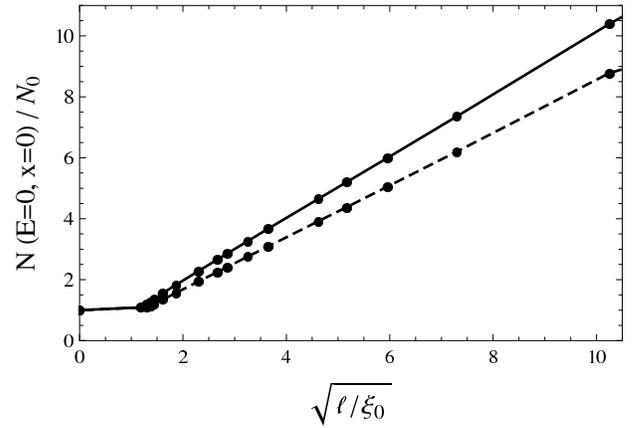}
\caption{\label{Fig6} Zero energy density of states at the surface as a function of $\sqrt{\frac{l}{\xi_{0}}}$ for two different orientations $\alpha=\frac{\pi}{4}$ (solid) and $\alpha=\frac{\pi}{8}$ (dashed). Temperature is $T=0.1 T_c$. }
\end{center}
\end{figure}

These results show that the influence of Born impurity scattering is much
stronger at the surface than in the bulk due to the presence of the 
Andreev bound states.
Their presence creates a larger number of available scattering channels, which
in turn leads to a stronger broadening of the Andreev bound states.
This self-consistent broadening can be illustrated by looking at the peak height
of the local density of states at zero energy $N(E=0)/N_0$. On the one hand
the peak height scales approximately with the inverse of the local quasiparticle
scattering rate:
\begin{eqnarray*}
\frac{N(E=0)}{N_0} &\sim& \frac{k_B T_c}{- \mbox{Im} \Sigma^G(E=0)}
\end{eqnarray*}
On the other hand the local quasiparticle scattering rate is determined
by the peak height via Eq.~(\ref{selfenergies}):
\begin{eqnarray*}
- \mbox{Im} \Sigma^G(E=0) &=& \frac{1}{2\tau} \frac{N(E=0)}{N_0} 
\end{eqnarray*}
Solving for the peak height leads to the expression
\begin{eqnarray*}
\frac{N(E=0)}{N_0} &\sim& \sqrt{2 \tau k_B T_c} \sim \sqrt{\frac{l}{\xi_0}}
\end{eqnarray*}
This result is in good agreement with the numerical result shown in
Fig.~\ref{Fig6}. It shows that the scaling
behavior of the peak height is $\sim \sqrt{l}$ instead of the $\sim l$
behavior one would have expected from bulk scattering, leading to a stronger
impurity effect near the surface.

\begin{figure}
\begin{center}
\includegraphics[width=0.45\textwidth]{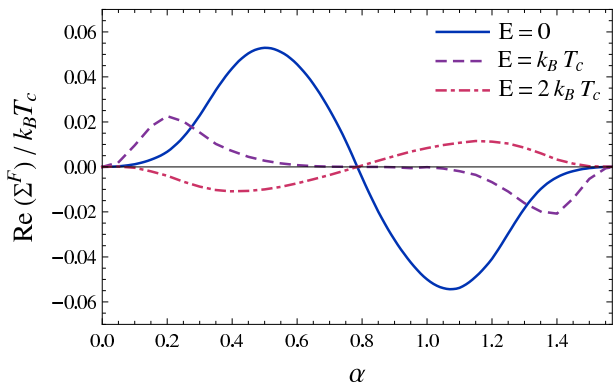}
\caption{\label{Re-Self-F-vs-alpha-x-0} (Color online)
Real part of self energy $\Sigma^{F}$ 
at the surface  vs. orientation angle $\alpha$ for different energies. 
The temperature is $T = 0.5 T_{c}$ and the mean free path $l = 2.7 \xi_{0}$.}
\end{center}
\end{figure}

\begin{figure}
\begin{center}
\includegraphics[width=0.33\textwidth, angle=270]{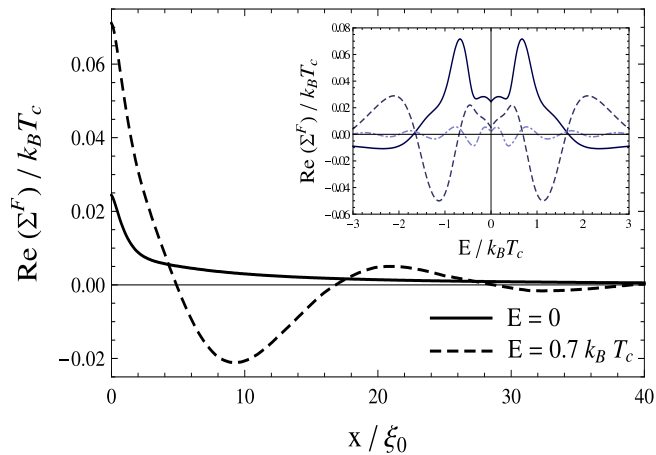}
\caption{\label{Re-Self-F-vs-x-alpha-pi-8-with-inset} (Color online)
Real part of the 
self energy $\Sigma^{F}$ vs. distance for two different energies $E$. The 
inset shows the energy dependence of this self energy for different
distances from the surface: $x/\xi_0=0$ (solid line),
$x/\xi_0=5$ (dashed line), and $x/\xi_0=20$ (dash-dotted line).
In both cases, the orientation is given by $\alpha= \frac{\pi}{8}$, 
the mean free path $l = 2.7 \xi_{0}$, and the temperature is $T = 0.1 T_{c}$.  }
\end{center}
\end{figure}

In an isotropic $s$-wave superconductor Anderson's theorem asserts that the
renormalization of the pair potential due to the anomalous self energy 
$\Sigma^F$ exactly
compensates the renormalization due to the normal self energy $\Sigma^G$,
such that the density of states and $T_c$ remain unaffected by
impurity scattering. This does not necessarily hold anymore in an
anisotropic superconductor, however \cite{Schopohl2}.
In the bulk of a $d$-wave superconductor the anomalous self energy 
$\Sigma^F$ is known to vanish. This is clear from 
Eq.~(\ref{selfenergiesF}), because the Fermi surface average leads
to cancellation due to the sign change of the $d$-wave. Ultimately,
this is the reason why nonmagnetic impurity scattering is much
more destructive to unconventional superconductors than to conventional
ones. It has not been noted before, however, that this vanishing of 
$\Sigma^F$ for $d$-wave superconductors
is not generally true anymore near the surface. Near the
surface translational invariance is broken, which makes trajectories
with different momenta $k_F$ inequivalent, because they experience
different pair potential landscapes. Except for special orientations
$\alpha$ of the $d$-wave with respect to the surface this leads
to finite values of the anomalous self energy $\Sigma^F$.
In Fig.~\ref{Re-Self-F-vs-alpha-x-0} we show the real part of 
$\Sigma^{F}$ at the surface as a function of the orientation angle 
$\alpha$ for different energies.
It can be seen that $\Sigma^{F}$ vanishes for integer multiples of
$\pi/4$. Fig.~\ref{Re-Self-F-vs-x-alpha-pi-8-with-inset} shows how
$\Sigma^{F}$ varies with the distance from the surface and energy (inset)
for $\alpha=\pi/8$ decreasing to zero in the bulk.

\section{\label{sec:3} Influence of an external magnetic field in the 
clean limit}

In this section we will discuss the influence of an external magnetic field
in the absence of impurity scattering. In particular we focus on the
case $\alpha = \frac{\pi}{4}$, where the spectral weight of the
Andreev bound states is strongest and $\kappa = 10$. This 
value of $\kappa$ is modest in comparison with $\kappa$ values of
hole doped high-$T_c$ cuprates, but may be relevant for some
low $T_c$ electron doped cuprates \cite{Fabrega}.
The influence of the anomalous Meissner effect becomes
more pronounced for small values of $\kappa$ and here we wish to illustrate
a peculiar effect that occurs in this range of parameters.

\begin{figure}
\begin{center}
\includegraphics[width=0.45\textwidth]{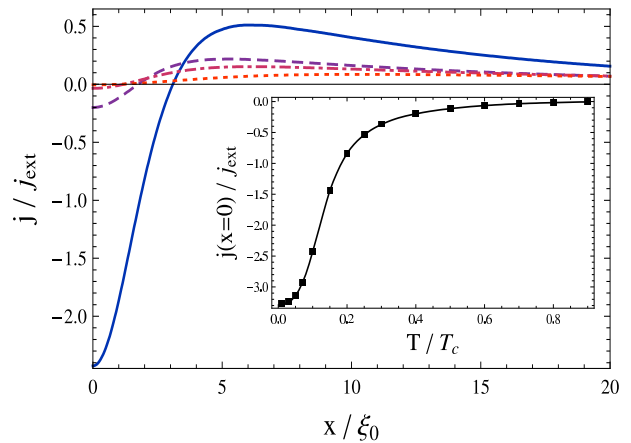}
\caption{\label{strom-pi-4-x-scale} (Color online)
Current density distribution for orientation
$\alpha = \frac{\pi}{4}$ at temperatures $T = 0.1 \, T_{c}$ (solid), 
$ 0.4 \, T_c$ (dashed), 
$ 0.7 \, T_c $ (dash-dotted), $ 0.9 \, T_c $ (dotted). 
The external magnetic field is $B_{ext} = 0.02 B_{c2}$. The inset shows the 
temperature dependence of the surface current density for orientation
$\alpha = \frac{\pi}{4}$ and the same value of the external magnetic field.
Here, the current density has been normalized to
$j_{ext}=\frac{c}{4\pi}\frac{B_{ext}}{\lambda_L}$.}
\end{center}
\end{figure}
 
In the following we have set the external magnetic field to $B_{ext} = 0.02 B_{c2}$. 
Fig.~\ref{strom-pi-4-x-scale} shows the current density for selected temperatures
as a function of the distance from the surface. 
In a distance up to $3 \xi_{0}$ from the surface (which is of the order of the 
spatial extension of the bound states), the current is flowing opposite to 
the screening current \cite{Fogelstroem,Walter}. This anomalous Meissner current 
persists throughout 
the full temperature range between $0.01 T_{c}$ and $0.9 T_{c}$, as demonstrated 
in the inset of Fig.~\ref{strom-pi-4-x-scale}. 
While it is nearly vanishing for temperatures close
to $T_{c}$, it saturates near zero temperature .

\begin{figure}
\begin{center}
\includegraphics[width=0.45\textwidth]{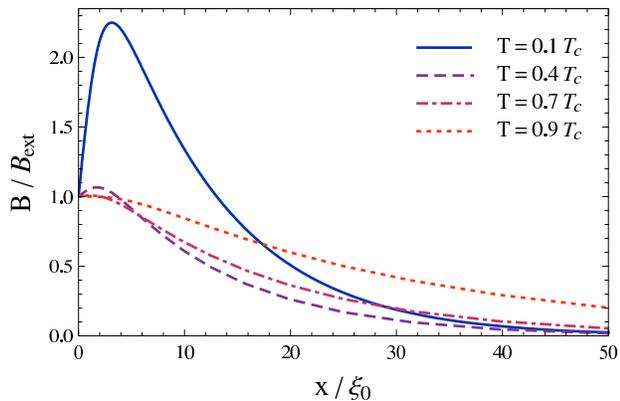}
\caption{\label{magfeld-pi-4-x-scale} (Color online)
Magnetic field as a function of
the distance from the surface for orientation
$\alpha = \frac{\pi}{4}$ and different temperatures. 
The external magnetic field is $B_{ext} = 0.02 B_{c2}$. }
\end{center}
\end{figure}

The magnetic field distributions resulting from the current distributions are shown in 
Fig.~\ref{magfeld-pi-4-x-scale}. With the anomalous Meissner current flowing, 
the magnetic field initially increases before the normal Meissner screening
sets in and eventually screens out the magnetic field exponentially. This 
initial increase occurs again up to a distance of $\sim 3 \xi_{0}$ from the 
surface.  With the value of $\kappa=10$ we have used here, 
the field increases by more than a factor of two relative to the external 
field. Qualitatively it is clear that this field increase becomes more
pronounced for smaller values of $\kappa$, because a smaller penetration
depth results in larger current densities, as seen from Eq.~(\ref{math-num-strom}).

\begin{figure}
\begin{center}
\includegraphics[width=0.45\textwidth]{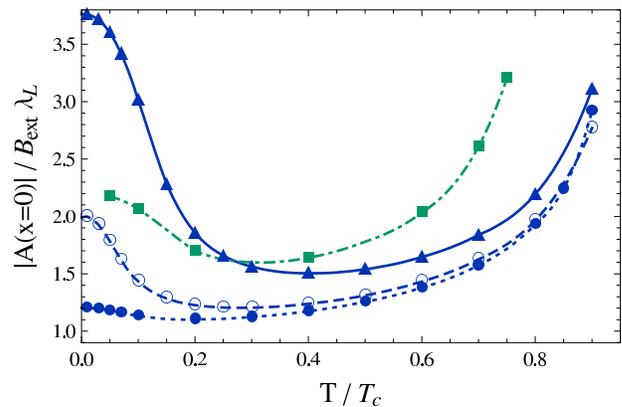}
\caption{\label{magveksurf-pi-4-TT-versch} (Color online)
Temperature dependence of the 
surface vector potential for orientation $\alpha = \frac{\pi}{4}$.
Results are shown for $\kappa=10$ and $B_{ext} = 0.02 B_{c2}$
(triangles), $\kappa=30$ and $B_{ext} = 0.006 B_{c2}$ (open circles),
$\kappa=63$ and $B_{ext} = 0.006 B_{c2}$ (filled circles).
For comparison, the squares show the result for $\kappa=10$ and
$B_{ext} = 0.02 B_{c2}$ including impurity scattering with
a mean free path of $l = 5.3 \xi_{0}$. Lines are guide to
the eye. }
\end{center}
\end{figure}
     
Fig.~\ref{magveksurf-pi-4-TT-versch} shows the modulus of the 
vector potential at the surface as a function of temperature. It can be
seen that the temperature dependence is nonmonotonous. The vector potential
increases both towards low temperatures as well as towards $T_c$.
The behavior near $T_c$ is easily understood from the temperature
dependence of the penetration depth, which diverges near $T_c$.
Since the vector potential is the integral of the magnetic field,
at a fixed external magnetic field we have to expect an increasing
vector potential with increasing penetration depth.
The increase of the surface vector potential towards low temperatures
has a different physical origin: it is directly related to the
anomalous Meissner effect and the field increase shown in
Fig.~\ref{magfeld-pi-4-x-scale}.

\begin{figure}
\begin{center}
\includegraphics[width=0.45\textwidth]{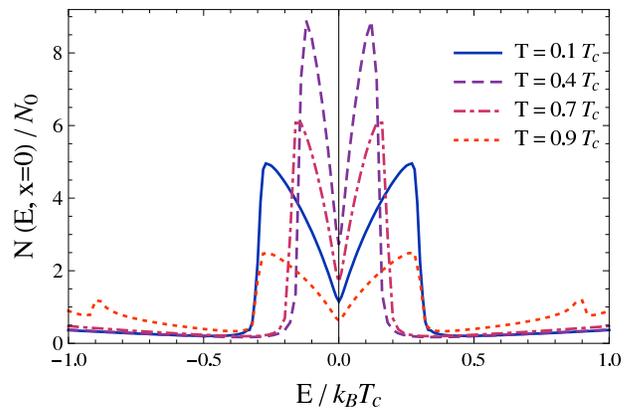}
\caption{\label{ndichte-pi-4-magfeld} (Color online)
Nonmonotonous splitting of the local 
density of states for orientation $\alpha = \frac{\pi}{4}$ at different
temperatures.
The magnetic field is $B = 0.02 B_{c2}$. }
\end{center}
\end{figure}

Since the vector potential is proportional to the superfluid velocity,
this nonmonotonous temperature dependence of the vector potential has a
direct influence on the size of the peak splitting in the local density of 
states\cite{Fogelstroem}, which we show in Fig.~\ref{ndichte-pi-4-magfeld}. 
It can be seen that the splitting is large both for low temperatures
and close to $T_c$. As a result also the peak height has a nonmonotonous
temperature dependence. The observation of such an increase of the peak 
splitting towards low temperatures could be an experimental signature
of the anomalous Meissner currents. It should be pointed out, however,
that this effect becomes less pronounced the larger the $\kappa$ value
of the material. This is shown in Fig.~\ref{magveksurf-pi-4-TT-versch}
for $\kappa=30$ and $\kappa=63$ as the open and solid circles,
respectively. For these higher values of
$\kappa$ the increase of the vector potential
towards low temperatures is gradually reduced. 
We also want to mention that impurity scattering gradually reduces 
this low temperature increase. The squares in 
Fig.~\ref{magveksurf-pi-4-TT-versch} show the behavior 
for $\kappa=10$ and a mean-free path of $l=5.3\xi_0$. 
The low temperature increase is reduced, while the increase
towards $T_c$ is shifted due to the reduction of the bulk $T_c$.

\begin{figure}
\begin{center}
\includegraphics[angle=270,width=0.45\textwidth]{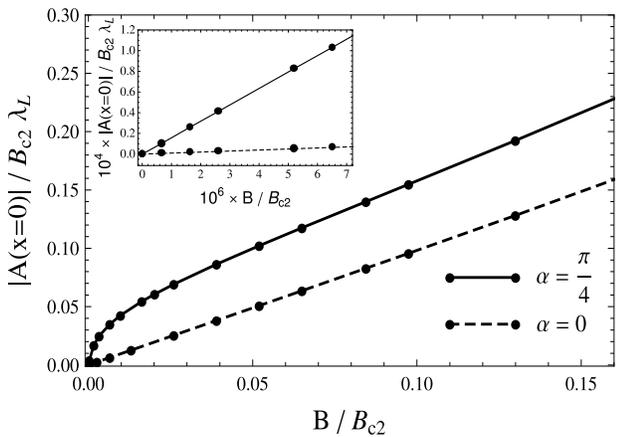}
\caption{\label{magveksurf-pi-4-Bext}
Surface vector potential as a function of the external magnetic field
$B/B_{c2}$ for temperature $T=0.1 T_c$ and $\kappa=10$. The solid line 
shows the result for $\alpha = \frac{\pi}{4}$ and the dashed line for 
$\alpha = 0$. In the inset the low field range is shown.
 }
\end{center}
\end{figure}

The presence of the surface Andreev bound states also has a
significant influence on the nonlinear Meissner effect.
We demonstrate this in Fig.~\ref{magveksurf-pi-4-Bext},
which shows the surface vector potential as a function of the 
external magnetic field for orientations $\alpha = \frac{\pi}{4}$ 
(solid line) and $\alpha = 0$ (dashed line). For $\alpha = 0$,
where surface Andreev bound states are absent,
the response is linear over a broad range of magnetic fields.
In contrast, for $\alpha = \frac{\pi}{4}$ sizeable nonlinear corrections
are visible, seen as a steep increase at low fields. In the low
field range of the order of $\sim 10^{-5} B_{c2}$, shown in the inset,
the response is linear in both cases with a significantly larger
slope at $\alpha = \frac{\pi}{4}$. We suppose that this surface
related effect may have an important influence on the nonlinear
Meissner effect in $d$-wave superconductors and intermodulation
distortion generated in high-$T_c$ microwave resonators 
\cite{Yip,Dahmimd,Oates}.

\section{\label{sec:4} Influence of impurity scattering on the peak splitting}

\begin{figure}
\begin{center}
\includegraphics[width=0.45\textwidth]{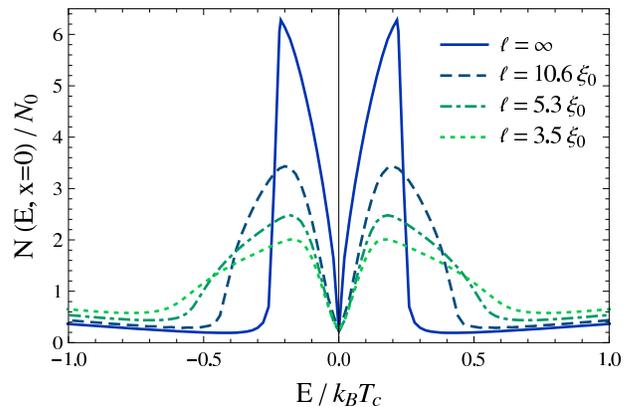}
\caption{\label{Dos-wall-pi-4-B-impurity-T-01} (Color online)
Local density of states at the 
surface for different impurity mean free paths in the presence of an external 
magnetic field $B = 0.006 B_{c2}$. The temperature is $ T=0.1 T_c$ and the 
orientation angle $\alpha=\frac{\pi}{4}$. }
\end{center}
\end{figure}

Having discussed the two limiting cases of impurity scattering without an external
magnetic field and the influence of a magnetic field without impurity scattering,
we turn now to a discussion of the combined effect of impurity scattering in
the presence of an external magnetic field. Naively, one might expect that
the impurity scattering will wash out the peak splitting. We will show below
that the situation is more complex, however. In the following we will
work with a value of $\kappa=63$, which is more realistic for hole doped 
high-$T_c$ cuprates.

\begin{figure}
\begin{center}
\includegraphics[width=0.45\textwidth]{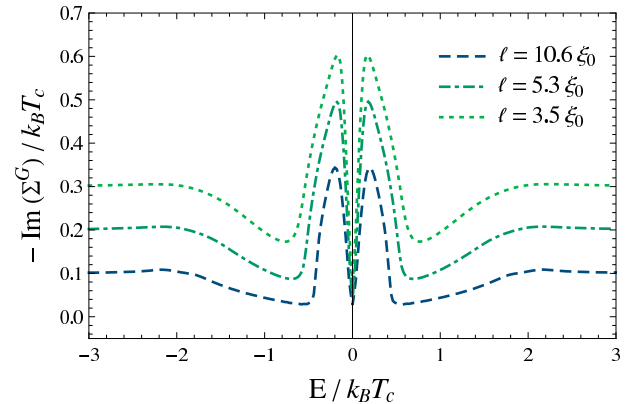}
\caption{\label{Im-Self-G-vs-E-with-B-Pi-4} (Color online)
Negative imaginary part of the self energy $\Sigma^{G}$ at the 
surface for different impurity concentrations in the presence of an external 
magnetic field $B = 0.006 B_{c2}$. The temperature is $ T=0.1 T_c$ and the 
orientation angle $\alpha=\frac{\pi}{4}$.}
\end{center}
\end{figure}

In Fig.~\ref{Dos-wall-pi-4-B-impurity-T-01} we demonstrate the influence
of impurity scattering on the peak splitting of the local density of
states at the surface. Here we have chosen a surface angle $\alpha=\frac{\pi}{4}$,
temperature $ T=0.1 T_c$, and an external magnetic field of $B = 0.006 B_{c2}$.
It can be seen that the peak height is strongly reduced and the peak width
grows with decreasing mean free path. However, the size of the peak splitting
remains almost unaffected. This peculiar effect can be understood from the
energy dependence of the negative imaginary part of the self energy $\Sigma^{G}$,
which is shown in Fig.~\ref{Im-Self-G-vs-E-with-B-Pi-4}. Here we can see
that the scattering rate also shows a splitting in energy, which is just
the mirror image of the splitting in the local density of states. It results
from the fact that due to the peak splitting the available phase space for 
scattering processes is strongly reduced at low energies. This in turn
means that the quasiparticle scattering rate at small energies remains small even when
the mean free path becomes small. This leads to a self-stabilization of
the peak splitting making it robust against impurity scattering.

\begin{figure}
\begin{center}
\includegraphics[width=0.45\textwidth]{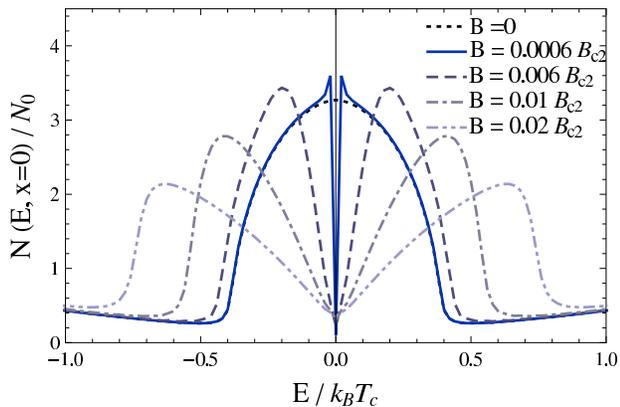}
\caption{\label{Fig16} (Color online) Local density of states at the surface 
for different external magnetic fields with mean free path $l = 10.6 \xi_{0}$. 
The temperature is $ T=0.1 T_c$ and the orientation angle $\alpha=\frac{\pi}{4}$.}
\end{center}
\end{figure}

In Fig.~\ref{Fig16} we show the peak splitting for a fixed mean free path 
$l=10.6 \xi_0$
and a series of external magnetic fields. When the magnetic field is increased,
the peak splitting does not evolve as a dip near zero energy, which gradually
becomes deeper, but instead the splitting opens up like a curtain. Note that
the frequency dependence at low energy is approximately
linear over an increasing energy
scale and the two peaks get a triangular lineshape.

\begin{figure}[t]
\begin{center}
\includegraphics[width=0.45\textwidth]{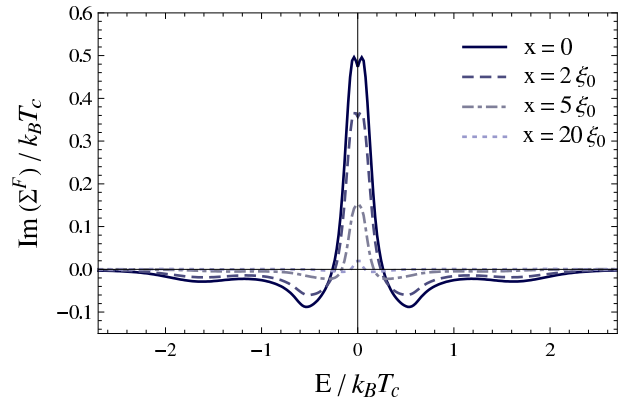}
\caption{\label{Im-Self-F-vs-E-with-B-Pi-4} (Color online) 
Imaginary part of the off-diagonal 
self energy with external magnetic field $B = 0.006 B_{c2}$ for orientation 
$\alpha=\frac{\pi}{4}$, temperature $T = 0.1 T_{c}$ and $l = 3.5 \xi_{0}$.}
\end{center}
\end{figure}

In the absence of an external magnetic field we have pointed out above
that the anomalous self energy $\Sigma^{F}$ does not vanish except
for certain highly symmetric angles like $\alpha=\pi/4$. In the
presence of an external magnetic field even for $\alpha=\pi/4$,
the anomalous self energy $\Sigma^{F}$ does not vanish anymore. 
This is shown in 
Fig.~\ref{Im-Self-F-vs-E-with-B-Pi-4}, where the imaginary part
of $\Sigma^{F}$ is shown as a function of energy for different distances
from the surface. The reason for this is that the special reflection
symmetry of the case $\alpha=\pi/4$ is broken now by the direction
of the current flow. 

\section{\label{sec:conclusion} Conclusions}

We have studied the influence of self-consistent Born impurity 
scattering on the
surface Andreev bound states in a $d$-wave superconductor. In
the absence of an external magnetic field Born impurity scattering
leads to a broadening of the zero energy peak in agreement with
previous work \cite{Poenicke,Tanaka}. We have shown, however, that
the effect of the impurities is much stronger at the surface than
in the bulk. Also, the renormalization of the pair potential
in general does not vanish anymore near the surface. 

In the presence of an external magnetic
field we have demonstrated that for small values of 
$\kappa\sim 10$ the zero energy peak splitting 
has a strongly nonmonotonous temperature dependence. 
For fixed external magnetic field the peak splitting
is larger at small and at high temperatures and has a minimum
in the intermediate temperature range. We have also shown that
the presence of Andreev bound states leads to significant nonlinear
corrections to the Meissner effect.
The range of $\kappa$ values in question is small compared with
hole doped high-$T_c$ cuprate materials. However, we suggest that these
effects may be observable in some electron-doped cuprates, which have
smaller $\kappa$ values due to a lower $T_c$ and larger carrier
densities \cite{Fabrega}. 

We have shown further, that the peak
splitting turns out to be quite robust against Born impurity 
scattering. This results from a self-stabilizing
effect of the peak splitting.

\acknowledgments
This work was supported by the Deutsche Forschungsgemeinschaft through
project No. DA~514/2-1. We thank S.~Graser and C.~Iniotakis for valuable discussions.

\end{document}